\documentstyle[preprint,aps,epsfig]{revtex}
\tightenlines
\begin{document}
\title{ON THE SCALAR POTENTIAL MODELS FROM  THE ISOSPECTRAL POTENTIAL CLASS}
\author{V. Gomes Lima$^{(a)},$ V. Silva Santos$^{(b)}$ and
R. de Lima Rodrigues$^{(c)}$\thanks{Permanent address:
Departamento de Ci\^encias Exatas e da Natureza, Universidade
Federal de Campina Grande, Cajazeiras - PB, 58.900-000 - Brazil,
e-mail: rafaelr@cbpf.br ou rafael@cfp.ufpb.br} \\
${}^{(a,c)}$ Centro Brasileiro de Pesquisas F\'\i sicas (CBPF)\\
Rua Dr. Xavier Sigaud, 150\\
 CEP 22290-180, Rio de Janeiro-RJ, Brazil\\
${}^{(a)}$ Departamento de F\'\i sica,
Universidade Federal Rural do Rio de Janeiro\\
Antiga Rodovia Rio-S\~ao Paulo Km 47, BR 465\\
CEP 23.890-000, Serop\'edica-RJ - Brazil\\
${}^{(b)}$ Departamento de Engenharia Civil,
Universidade Federal de Campina Grande\\
CEP 58.109-970, Campina Grande - PB - Brazil}

\maketitle

\begin{abstract}
Static field classical configurations in (1+1)-dimensions
for  new  non-linear potential models are investigated
from  an isospectral potential class and the concept of bosonic
zero- mode solution.
One of the models here considered has a static non-topological configuration
with a single vacuum state that  breaks supersymmetry.
\end{abstract}


\newpage

\section{Introduction}

Rencently, static classical configurations that are exact solution
of the equation of motion in scalar potential model have drawn much attention
mainly because they have found interesting  applications
as a solitary wave \cite{Rajaraman82}.
There are several methods to build up soliton (kink)
solutions associated with some nonlinear differential
equations \cite{Drazin}.
The kink of a scalar field theory  is a static, non-singular,
{\it classically stable} and finite-localized energy solution of the
equation of motion \cite{Bala}. When $\phi(x,t)=f(x-vt),$ a kink solution
is called solitary wave.
The role of classical solutions in quantum field theories have recently
been overviewed \cite{Weinberg92}. There, one can see how a quantum
field theory has topological and non-topological  soliton solutions
in higher spatial dimensions. In the literature, there is
a surprising number of scalar potential models in higher spatial dimensions
with exactly solvable equations of motion. Also, a
variety of classical finite-energy static solutions are known.

The connection between supersymmetry in quantum mechanics (SUSY
QM) \cite{Witten,La,Nieto84,Fred,RVV01} and the topological and
non-topological solitons in terms of a scalar potential (for the
case of a single field) has been discussed in the literature
\cite{K87,R95,Kuls,Sukumar86,R99}. The connection between SUSY QM
associated with two-component eigenfunctions \cite{RVV01} and the
topological solitons in terms of two coupled  scalar  fields  have
already been considered in \cite{RPV98}. Recently, the
reconstruction of 2-dimensional scalar field potential models has
been considered starting from the Morse and the Scarf II
hyperbolic potential, and quantum corrections to the solitonic
sectors of both potentials are pointed out \cite{GN}.

In the present  work, from a non-polynomial potential in
supersymmetric quantum mechanics, we shall propose  new potentials
as functions of one single real scalar field. Indeed, in (1+1)
dimensions, non-polynomial interactions are not really harmful,
from the point of  view of renormalisation. Actually, non-linear
$\sigma-$models are the most representative class of
renormalisable models in 2D, and their interactions are described
by non-polynomial functions. We consider the interesting program
of proposing new potential models in 1+1 dimensions, whose
essential point is associated with the translational invariance of
the static field configuration \cite{K87}. Here, the aim is to
find field potential models via SUSY QM considering a
one-dimensional quantum mechanical isospectral potential class
such that the corresponding fluctuation operator sets a
Schr\"odinger-like eigenvalue problem as a stability equation
exactly solvable  for the $\phi^4-$model. However, the
corresponding field potential cannot be put in a closed form.
Indeed, we show that in such a procedure there appears a new
non-polynomial field potential model with infinite-energy static
configuration, so that such a new potential may not be considered
as a well-defined theoretical field potential model.

Actually, we
shall bring two constructions of classical solutions.
In the first case,  we obtain
the well-known topological kink, whereas in the second case
we obtain a non-topological static configuration
which is not a kink. In the second case the new
static field configuration is not a smooth function
over all the the spatial-time.

We also propose some pictures of the field potential models
(figures I and III) and the static classical field configurations
(figures II and IV).

\section{Static Classical Configuration}

Consider the Lagrangian density for a single scalar
field, $\phi(x,t),$ in (1+1)-dimensions, in natural system  $(c=1=\hbar)$,
given by

\begin{equation}
{\cal L}\left(\phi, \partial_{\mu}\phi\right) = \frac{1}{2}
\partial_{\mu}\phi\partial^{\mu}\phi - V\left(\phi\right)
\label{E1},
\end{equation}
where  $V(\phi)$ is any positive semi-definite
function of $\phi$, which must have at least two zeros for kinks to exist.
It represents a well-behaved potential energy.
However, as it will be shown below, we have found a new potential
which is exactly solvable exactly in the context of the classical
theory in (1+1)-D that  not a kink.

The field equation for a static
classical configuration, $\phi =
\phi_c\left(x\right),$ becomes

\begin{equation}
\label{E2}
-\frac{d^2}{dx^2} \phi_c\left(x\right) +
\frac{d}{d\phi_c}V\left(\phi_c\right)
=0, \qquad \dot\phi_c = 0,
\end{equation}
with the following boundary conditions:
$\phi_c(x)\rightarrow \phi_{vacuum}(x)$
as $x\rightarrow \pm \infty.$
Since the potential is positive, it can be written as

\begin{equation}
\label{E4}
V(\phi) = \frac{1}{2} U^2(\phi),
\end{equation}
giving the well-known Bogomol'nyi condition,

\begin{equation}
\label{E5}
\frac{d\phi}{dx} = \pm U(\phi)
\end{equation}
where the solutions with the plus and minus signs represent two
configurations.

\section{Stability Equation and New Potentials}

The classical stability of the static solution is investigated
by considering small perturbations around it,

\begin{equation}
\label{E11}
\phi(x,t) = \phi_c(x) + \eta (x,t),
\end{equation}
where we expand the fluctuations in terms of the normal modes,
\begin{equation}
\label{E12}
\eta (x,t) = \sum_n \epsilon_n \eta_n (x) e^{i\omega_n t};
\end{equation}
the $\epsilon_n'$s are so chosen that $\eta_n (x)$ are real.
A localised classical configuration is said to be dynamically stable if
the fluctuation does not destroy it.
The equation of motion becomes a Schr\"odinger-like equation, viz.,

\begin{equation}
\label{E13}
F\eta_n (x)
= \omega_n{^2}\eta_n (x), \quad F=-\frac{d^2}{dx^2} +
V^{\prime \prime}(\phi_c).
\end{equation}

According to (\ref{E4}), one  obtains the supersymmetric form,
\cite{R99}

\begin{equation}
\label{E14}
V^{\prime\prime}(\phi_c) =
{U^\prime}^2 (\phi_c) + U(\phi_c)U^{\prime \prime}(\phi_c),
\end{equation}
where the primes stand for a second derivative with respect to the
argument.

If the normal modes of (\ref{E13}) satisfy
$\omega_n{^2} \geq 0$  the
stability of the Schr\"odinger-like equation is ensured.
Now, we are able to implement a method that provides a new potential
from  the potential term that appears in the fluctuation operator.

Next, we consider the following generalized isospectral potential
as being the potential term (\ref{E14}) for the fluctuation operator:

\begin{eqnarray}
\label{NP}
V(x;\alpha,\beta) &=& m^2\left[3tanh^2\left(\frac{m}{\sqrt 2}x \right) -1
\right]\nonumber\\
{}&+&
2m^2\beta \left[sech^4\left(\frac{m}{\sqrt 2}x \right)
\left(2tanh\left(\frac{m}{\sqrt 2}x \right)+
\frac{\beta}{2}sech^4\left(\frac{m}{\sqrt 2}x\right)
\chi \right)\chi\right],\nonumber\\
\chi &=& \chi (x;\alpha,\beta)=\left\{\alpha+
\beta\left[tanh\left(\frac{m}{\sqrt 2}x \right)
- \frac 13tanh^{3}\left(\frac{m}{\sqrt 2}x\right)\right]\right\}^{-1},
\end{eqnarray}
where $\alpha$ and $\beta$ are constant parameters.
This non-polynomial potential satisfies the condition $\omega_n{^2} \geq 0,$
and the
ground state associated to the zero mode ($\omega_0^2=0$) is given by

\begin{equation}
\label{GS}
\eta^{(0)} (x;\alpha,\beta)
=N\chi(x) sech^{2}\left(\frac{m}{\sqrt 2}x \right),
\end{equation}
where $N$ is the normalization constant. Note that, if
$|\alpha|>\frac{2\beta}{3},$ the eigenfunction of the ground state
is non-singular. This ground state was independently found by
two authors \cite{K87,R95} for $\beta=\frac 12\sqrt{\frac{3m}{\sqrt{2}}}.$

It is well-known that the bosonic zero-mode eigenfunction
of the stability equation is related with the kink by

\begin{equation}
\label{MZ}
\eta^{(0)} (x;\alpha,\beta)
\propto\frac{d}{dx}\phi_c(x),
\end{equation}
so that, a priori, we may find the static classical configuration
by a first integration. Therefore, the potential

\begin{equation}
V(\phi;\alpha,\beta)=\frac 12\left(\frac{d\phi}{dx}\right)^2
\end{equation}
yields a class of scalar potentials, $V(\phi)=V(\phi;\alpha,\beta),$
which have exact solutions.

There, we can find various static field configurations; however,
let us consider here only two cases.

Case (i): $\beta=0$ and $\alpha=2N\sqrt{\frac{\sqrt{2}}{3m}}$

In this case, $\chi (x)\rightarrow \frac{1}{\alpha},$  so that from
 (\ref{NP}) and (\ref{GS}), we obtain the following bosonic zero-mode
solution:

\begin{equation}
\label{MZPD}
\eta^{(0)} (x)=\frac 12\sqrt{\frac{3m}{\sqrt{2}}}
sech^{2}\left(\frac{m}{\sqrt 2}x \right),
\end{equation}
and the non-polynomial potential becomes

\begin{equation}
V_{1-}(x)=m^2\left[3tanh^2\left(\frac{m}{\sqrt{2}}x\right)
-1\right].
\end{equation}
Notice that the two Schr\"odinger-like fluctuation operators
associated with both $V(x;\alpha,\beta)$ and $V(x)$ non-polynomial
potentials are positive semi-definite and completly isospectrals.
However, their factorization has  been implemented from distinct
superpotentials. Indeed, while the Ricatti equation,

\begin{equation}
V_{1-}(x) = W_1^2(x) + W_1^\prime(x),
\end{equation}
where $W^\prime(x)=\frac{d}{dx}W(x),$ has a particular solution
given by
\begin{equation}
\label{SP1}
W_1(x)=-\sqrt{2}mtgh\left(\frac{m}{\sqrt{2}}x\right),
\end{equation}
provides

\begin{equation}
\label{EEF1}
\eta_{1-}^{(0)}(x)=e^{\int W_1(y)dy}=\frac 12\sqrt{\frac{3m}{\sqrt{2}}}
sech^{2}\left(\frac{m}{\sqrt 2}x \right)=\eta^{(0)}(x),
\end{equation}
with $\eta^{(0)}(x)$ given by Eq. (\ref{MZPD}),
the Ricatti equation
 $$
V_{0-}=V(x;\alpha,\beta) = W_0^2 (x) + W_0^\prime(x)
$$
has a particular solution given by
$ W_0(x)=\frac{d}{dx}\ell n\left(\eta_0(x;\alpha,\beta)\right).$

By substituting Eq. (\ref{MZPD}) in Eq. (\ref{MZ}), we get the
well-known kink of the double-well potential,

\begin{equation}
\label{K}
\phi_k (x)
= \frac{m}{\sqrt\lambda}tanh\left(\frac{m}{\sqrt 2}x \right).
\end{equation}

When we express the position coordinate in terms of the
kink, i.e. $x=x(\phi_k),$ we find the $\phi^4-$potential model with
spontaneously broken symmetry in scalar field theory, viz.,

\begin{equation}
\label{PD}
V(\phi) = \frac{\lambda}{4}\left(\phi^2 - \frac{m^2}{\lambda}\right)^2.
\end{equation}
The mass of the kink is finite, but, in the next case, we obtain
an undefined kink mass. Such a classical configuration  cannot
represents a stable particle. The pictures of the potential
(\ref{PD}) and of the kink (\ref{K}) are in Figures I and II. Note
that both are smooth functions of the field and the spatial
coordinate, respectively.

Case (ii): $\alpha=0.$

In this case, from (\ref{NP}), (\ref{GS}) and (\ref{MZ}), we
obtain the following non-polynomial potential in the singularity
region associated to the ground state given by Eq. (10):

\begin{equation}
\label{E21}
\tilde{V}(\phi)=\frac{\lambda}{2}\left(1+3e^{-\gamma\phi}\right)
\left(1-\frac{2}{3}e^{\gamma\phi}\right)^2,
\end{equation}
where $\gamma=\frac{2}{\sqrt 3}$ and $\lambda$ is
a dimensionless constant.

The vacuum state, $\phi_{v},$ is given by
\begin{equation}
\label{E22}
\phi_{v} = \frac{\sqrt 3}{2}\ell n({\frac 32}).
\end{equation}
This non-polynomial potential does not have the discrete symmetry,
$\phi\rightarrow-\phi,$
and there exists only one vacuum state, so that it is non-topological.

From (\ref{E4}), (\ref{E5}) and (\ref{E21}), for the minus sign of the
Bogomol'nyi condition and the
coupling constant $\lambda\neq 0$, the static classical
configuration has the following explicit form

\begin{equation}
\label{E23}
\phi_c (x) =\frac{\sqrt 3}{2}\ell n
\left(\frac{tanh^{2}\left(\sqrt{\lambda}x\right)}
{1-\frac{1}{3}tanh^{2}\left(\sqrt{\lambda}x\right)}\right),
\end{equation}
where the integration constant is taken to be zero.
Note that this static configuration satisfies Eq. (\ref{MZ})
and the following boundary conditions $\phi_c(x)\rightarrow\phi_{v}$
as $x\rightarrow\pm\infty.$
The pictures of the non-polynomial potential and the static solution are in
Figs. II and IV. However, in Fig. IV, we have plotted this static field
configuration only for the region in that it has a singularity
in the origin.

The energy density for the static solution for the non-polynomial
potential is given by

\begin{equation}
\label{E24}
E(x) \propto\left\{ sinh^2\left(\sqrt{2}mx\right)
\left(1-\frac 13 tanh^2\left(\frac{m}{\sqrt 2}x \right)\right)^2
\right\}^{-1},
\end{equation}
which yields  an undefined total energy or classical mass. Here,
we have used the explicit relations between the static
configuration and one-dimensional spatial coordinate.

\section{Broken SUSY QM}

In this section, we consider the fluctuation operator
in the context of supersymmetry in Quantum Mechanics (SUSY QM), where
the supersymmetric partners are build up from the stability equation.
In $N=2-$SUSY, we define the following first order differential
operators:

\begin{equation}
\label{EA}
A_2^\pm =  \pm\frac{d}{dx} + W_2(x),\qquad
A^+_2 = \left(A^-_2\right)^\dagger.
\end{equation}
The fluctuation operator for the bosonic sector is given by

\begin{equation}
\label{E26}
F_{2-} \equiv A^+_2 A^-_2 = -\frac{d^2}{dx^2}+V_{2-}(x),
\quad V_{2-}(x)=\tilde{V}^{\prime \prime}_{|\phi=\phi_c},
\end{equation}
so that in terms of the superpotential we obtain the following
nonlinear first order differential equation
\begin{equation}
\label{E27}
 V_{2-} (x) = W_2^2 (x) + W_2^\prime(x)\equiv V(x;0, \beta),
\end{equation}
where the prime means a derivative with respect to $x.$

The superpotential that solve this Riccati equation for the
non-polynomial potential has the following explicit form:

\begin{equation}
\label{ESP}
W_2(x)=-\frac{m}{\sqrt 2}
\frac{tanh^4\left(\frac{m}{\sqrt 2}x \right)+3}
{tanh\left(\frac{m}{\sqrt 2}x \right)
[3-tanh^2\left(\frac{m}{\sqrt 2}x \right)]}
\end{equation}
Note that this particular solution to the Riccati differential
equation has the following asymptotic behavior:
$W_2(x)\rightarrow-\sqrt{2}m$ as $x\rightarrow\infty$ and
$W_2(x)\rightarrow \sqrt{2}m$ as $x\rightarrow -\infty.$ The
supersymmetric partner of $F_{2-}$ is given by

\begin{eqnarray}
\label{E28}
F_{2+} &&= A^-_2 A^+_2= -\frac{d^2}{dx^2} + V_{2+} (x) \nonumber\\
V_{2+} (x) &&= W_2^2(x) - W_2^{\prime}(x) =
m^2\left\{1 +tanh^2\left(\frac{m}{\sqrt 2}x \right)\right\}.
\end{eqnarray}
These fluctuation operators  are isospectral and consist of a pair
Schr\"odinger-like Hamiltonians of Witten's model of broken SUSY
\cite{Witten}. Note that the shape invariance condition
\cite{La,Fred} is not
satisfied for $V_{\pm}$ given by Eqs. (\ref{E27}) and (\ref{E28}),
i.e. $V_{2+}(x;a_2)\neq V_{2-}(x;a_1)+R,$ where $a_1, a_2$ and  $R$ are
constants.

The eigenvalue equations for the supersymmetric partners $F_\mp$ are
given by

\begin{equation}
\label{E30}
F_{2\pm}\eta^{(n)}_{2\pm}(x) = \omega^{(n)}_{2\pm}\eta^{(n)}_{2\pm}{(x)},\quad
\omega_{2-}^{(n)}=\omega_n^2, \quad\omega_0^2=0,
\end{equation}
for which, in general, $F_{2-}$ may have as eigenstates the
well-known normal modes. However, when $\alpha=0,$ the bosonic
zero-mode $(\omega_0^2=0)$ satisfies the annihilation condition

\begin{equation}
\label{E31}
A^-_2 \eta^{(0)}_{2-} = 0\Rightarrow\eta^{(0)}_{2-} (x)
\propto\frac{1}{{sinh\left(\sqrt{2}mx\right)
\left(2+sech^2\left(\frac{m}{\sqrt 2}x \right)\right)}}
=\eta^{(0)}(x;0,\beta).
\end{equation}
This eigensolution is not normalizable, so that the fluctuation
operator for the bosonic sector does not have a zero-mode. In this
case, the integral $\int_{-\infty}^{+\infty}
\left(\eta^{(0)}_{2-}(x)\right)^2dx$ is undefined. This result is
in agreement with Eq. (\ref{GS}), for $\alpha = 0.$ Furthermore,
 bosonic zero mode satisfies
$\eta^{(0)}_{2-}(x)= \frac{d}{dx}\phi_c (x).$ The fermionic sector
fluctuation operator $F_{2+}$ does not  have  zero-modes
because $\eta^{(0)}_{2+},$

\begin{equation}
\label{E32}
A^+_2 \eta^{(0)}_{2+} = 0\Rightarrow\eta^{(0)}_{2+} \propto
sinh\left(\sqrt{2}mx\right)\left[2+
sech^2\left(\frac{m}{\sqrt 2}x \right)\right]
\end{equation}
because it is not normalizable. In this case we have broken SUSY.
Indeed, it is easy to see that
$$
\int_{-\infty}^{+\infty}
\left(\eta^{(0)}_{2+}(x)\right)^2dx\rightarrow\infty.
$$
The eigenvalues $\omega_{\pm}$ and eigenfunctions $\eta^{(n)}_{2\pm}$
can be exactly solved in a similar way for a general potential.
All eigenvalues of $F_{2+}$ are eigenvalues of $F_{2-},$ i.e.
 $\omega_{2-}^{(n)}=\omega_{2+}^{(n)}>0,$
so that the ground state and the excited state
of both $F_{2\pm}$ have energy different from zero.

\section{Conclusion}

In this work, we investigate the classical stability of
a new isospectral non-polynomial potential model with
static classical configuration which solve exactly
the equation of motion.

Indeed, the classical finite-energy static solutions appear in field
theoretical models
with spontaneously broken symmetry (SUSY), for example,
in the double-well potential given by Eq. (\ref{PD}).
It is well-known that the double-well potential model
which has two zeros corresponding to the vacuum states
$\phi_1$ and $\phi_2$. In this case, the topological kink interpolates
smoothly and monotonically between $\phi_1$ and $\phi_2$, according to
Figs. I and II. But, in
our non-polynomial scalar potential one
builds a static, infinite-energy and classically
singular field configuration, which is in
a non-topological sector. Indeed, we have found a new potential
model  given by Eq. (\ref{E21}) that is solved exactly
in the context of classical theory in (1+1)-dimensions.

In conclusion, we found $V_{1-}$ and $V_{1+}=V_{2+}$ is a
supersymmetric potential pair with unbroken SUSY, so that $F_{1-}$
refers to the bosonic sector of the SUSY fluctuation operator
$F_{SUSY},$ while $F_{1+}$ is the fermionic sector of $F_{SUSY}$.
In this case, $\eta^{(0)}_{1-}$ then becomes the unique
normalizable eigenfunction  of the $F_{SUSY}$ corresponding to the
zero mode of the ground state. On the other hand, the spectra to
$F_{2\pm}$ are identical and in this case, there are no zero mode
for the ground state i.e. $\omega_-=\omega_+>0,$ thus one has
broken SUSY.

Therefore, our non-polynomial potential does not have the reflection
symmetry $\phi\rightarrow -\phi,$ with a stability equation so that it does
not lead to either a bosonic zero mode or its supersymmetric partner
because both eigenfunctions are non-normalizable with SUSY broken. Thus, the
 scheme above for proposing new field potential models is not always
physically  acceptable, because it may lead to infinite energy
configuration.

\vskip 1.0 true cm
\noindent{{\large\bf Acknowledgments}}

RLR wishes to thanks the  Dr. A. N. Vaidya of the Instituto de
F\'\i sica of the UFRJ for the incentive and the staff of CBPF and
DCEN-CFP-UFPB to facilities.  This research was supported in part
by CNPq (Brazilian Research Agency). RLR and VGL are grateful to
Dr. J. A. Helayel-Neto for his hospitality and fruitful
discussions.

\newpage


\unitlength=1cm
\begin{figure}[tbp]
\centering
\begin{picture}(10,1)
\epsfig{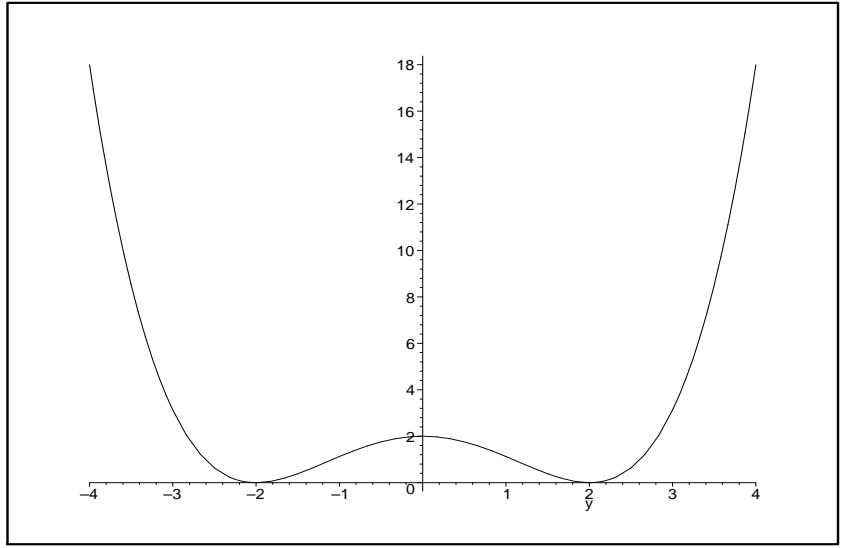}
\end{picture}
\vspace{7.5cm}
\caption{ Double-well potential
$V(y)= \frac{1}{8}\left(y^2-4\right )^2,$ for $\lambda=\frac 14,
\quad m=1.$ }
\end{figure}


\unitlength=1cm
\begin{figure}[tbp]
\centering
\begin{picture}(10,1)
\epsfig{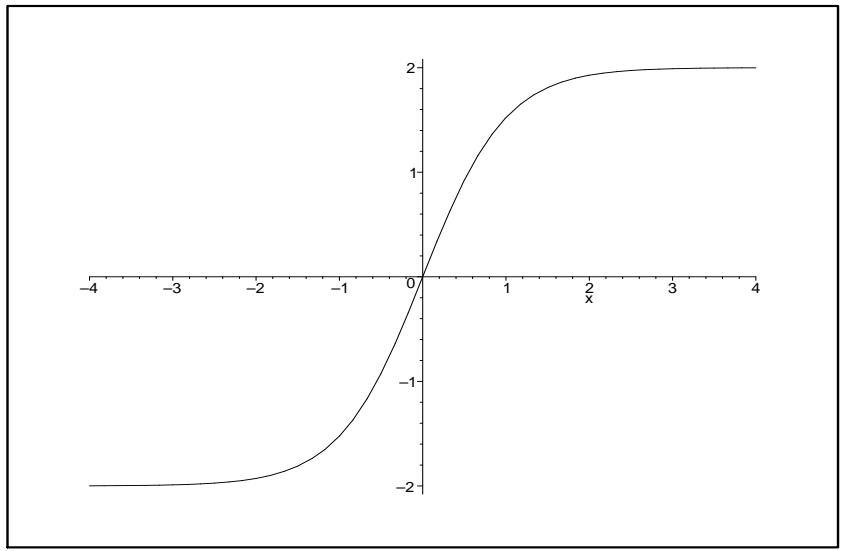}
\end{picture}
\vspace{7.5cm}
\caption{The Kink $y=\phi(x)$ of the double-well potential, for
$\lambda=\frac 14, \quad m=1.$}
\end{figure}

\newpage

\unitlength=1cm
\begin{figure}[tbp]
\centering
\begin{picture}(10,1)
\epsfig{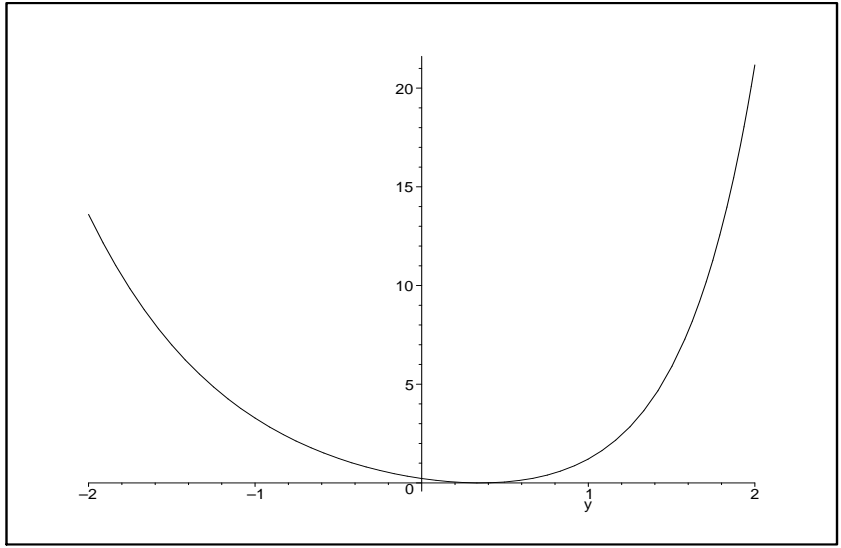}
\end{picture}
\vspace{7.5cm} \caption{ Non-polynomial potential, for
$\lambda=1,\quad m=1.$ }
\end{figure}


\unitlength=1cm
\begin{figure}[tbp]
\centering
\begin{picture}(10,1)
\epsfig{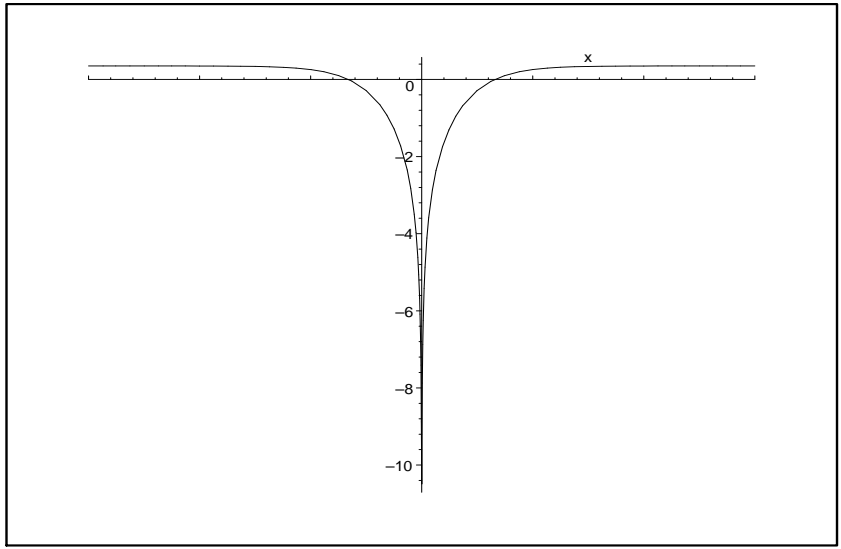}
\end{picture}
\vspace{7.5cm} \caption{ Static classical configuration associated
to the non-polynomial potential, for
$\lambda=1, \quad m=1.$ }
\end{figure}

\end{document}